\documentclass[12pt]{article}
\makeatletter

\@addtoreset{equation}{section}
\makeatother

\usepackage[nosort]{cite}

\begin{document}

\pagestyle{plain}
\setcounter{page}{1}
\begin{titlepage}

\rightline{\tt hep-th/0205283}

\vskip .1 cm

\rightline{\small{\tt TU-656}}

\begin{center}

\vskip 2 cm

{\LARGE\bfseries BPS Solutions of Noncommutative Gauge Theories 
in Four and Eight Dimensions}

\vskip 2cm
{\large Yoshiki Hiraoka}

\vskip 1.2cm

{\it Department of Physics, Tohoku University}

{\it Sendai 980-8578, JAPAN}

\vskip .7cm
{\tt hiraoka@tuhep.phys.tohoku.ac.jp}

\vskip 1.5cm

{\bf Abstract}
\end{center}

\noindent

We study the 1/4 BPS equations in the eight dimensional noncommutative
Yang-Mills theory found by Bak, Lee and Park.  
We explicitly construct some solutions of the 1/4 BPS equations 
using the noncommutative version of the ADHM-like construction in 
eight dimensions. 
From the calculation of topological charges, 
we show that our solutions can be interpreted 
as the bound states of the $D0$-$D4$-$D8$ with a $B$-field.
We also discuss the structure of the moduli space of 
the 1/4 BPS solutions and determine the metric of the moduli space 
of the $U(2)$ one-instanton in four and eight dimensions.

\end{titlepage}

\newpage


\section{Introduction}

Noncommutative geometry has played an important part 
in the study of string/M-theory~\cite{cds}.
In particular, $D$-branes with a constant NS $B$-field are of interest 
in the context of understanding 
the non-perturbative aspects of string theory.
The effective world-volume field theory on $D$-branes with a $B$-field 
turns out to be the noncommutative Yang-Mills theory~\cite{sw}, 
which has an interesting feature 
that the singularity of the instanton moduli space 
is naturally resolved~\cite{ns}. 

Four dimensional $U(N)$ $k$-instanton is realized as $k$ $D0$-branes 
within $N$ $D4$-branes in type IIA string theory.
When we turn on a self-dual constant $B$-field  
which preserves 1/4 of supersymmetries,  
the instanton moduli space is resolved and the $D0$-branes 
cannot escape from the $D4$-branes.
From the viewpoint of the $D0$-brane theory, 
the Higgs branch of the moduli space coincides with the moduli space 
of instantons and the self-dual $B$-field corresponds to  
the Fayet-Iliopoulos (FI) parameters.
If the FI parameters are non-zero, the $D0$-$D4$ system can not enter the 
Coulomb branch through the small instanton singularity.

It is also of interest to generalize the above system 
to higher dimensions in the context of both $D$-brane dynamics 
and the world-volume theories. 
The bound states of the $D0$-$D6$ and the $D0$-$D8$ with a $B$-field 
were investigated by several authors
~\cite{ohtan, cimm, park, witten, sato, fio, ohta, hio, kly3, pt, hiraoka}.  
These systems are equivalent by T-duality to the rotated branes at angles.
It has been shown that the $D0$-$D6$ system can form bound states only 
if the value of the $B$-field are taken appropriately  
and that in the $D0$-$D8$ system there are three cases preserving 
respectively 1/16, 1/8 and 3/16 of supersymmetries.  

These studies reduce to finding the solutions of 
the higher dimensional analogue 
of ``self-duality'' equations found in~\cite{cdfn, ward, hull}, 
which are the first order linear relations 
amongst components of the field strength.   
The above three cases of the $D0$-$D8$ systems correspond to 
the subgroup $Spin(7), SU(4)$ and $Sp(2)$ of the eight dimensional 
rotation group $SO(8)$ respectively. 
Recently all possible ``self-duality'' equations in the higher dimensional 
Yang-Mills theories have been classified in~\cite{blp}  
and it is shown that there are many kinds of the BPS equations 
which preserve $1/16, 2/16,\ldots, 6/16$ of supersymmetries. 
These include three new cases besides the above three ones.
In this paper, we will focus on the case of the 1/4 supersymmetries,  
construct some explicit solutions of the corresponding BPS equations 
and give their $D$-brane interpretations.

This paper is organized as follows. 
In section 2, we review the BPS equations in eight dimensions which 
have been derived in~\cite{blp}.
In section 3, 
we construct some solutions of the 1/4 BPS equations in eight dimensions
using the noncommutative version of the ADHM-like construction in eight 
dimensions.   
We show that our solutions can be interpreted as the bound states of the 
$D0$-$D4$-$D8$ with a $B$-field 
from the calculation of their topological charges. 
We also discuss the structure of the moduli space of 
the 1/4 BPS solutions and determine the metric of the 
moduli space of the $U(2)$ one-instanton in four and eight dimensions.
The final section is devoted to discussions.


\section{BPS equations in eight dimensions}

In this section, we briefly review the results of~\cite{blp}, in which  
all possible BPS equations in six and eight dimensional 
Yang-Mills theories have been classified. 
The authors of~\cite{blp} have investigated the higher dimensional analogue 
of ``self-duality'' equations, 
which are the linear relations amongst components of the 
field strength with the constant 4-form tensor $T_{abcd}$\,,
\begin{equation}
  F_{ab}+\frac{1}{2}T_{abcd}F_{ad}=0\,.\label{eq:2.1}
\end{equation}
This is a natural generalization of the 
four dimensional self-duality equation,
\begin{equation}
  F_{ab}+\frac{1}{2}\epsilon_{abcd}F_{ad}=0\,.
\end{equation}
When this equation holds, 
the equations of motion $D_aF_{ab}=0$ 
are automatically satisfied due to the Jacobi identity. 
We investigate only the eight dimensional case.
The results of~\cite{blp} are written as follows:

\begin{itemize}
\item The 1/16 BPS equations 
\begin{eqnarray}
& & F_{12}+F_{34}+F_{56} \pm F_{78}=0\,,\nonumber\\
& & F_{13}+F_{42}+F_{57} \pm F_{86}=0\,,\nonumber\\
& & F_{14}+F_{23}+F_{76} \pm F_{85}=0\,,\nonumber\\
& & F_{15}+F_{62}+F_{73} \pm F_{48}=0\,,\label{eq:2.2}\\
& & F_{16}+F_{25}+F_{47} \pm F_{38}=0\,,\nonumber\\
& & F_{17}+F_{35}+F_{64} \pm F_{82}=0\,,\nonumber\\
& & \pm F_{18}+F_{27}+F_{63} + F_{54}=0\,.\nonumber
\end{eqnarray}
These can be expressed compactly 
using the structure constants $C_{ijk}$ of the octonion,
\begin{equation}
  F_{i8}\pm \frac{1}{2}C_{ijk}F_{jk}=0\,,\label{eq:2.3}
\end{equation}
where $i,j,k=1,\cdots,7$.
Some solutions of these equations were constructed in~\cite{fn, ky}.

\item The 2/16 BPS equations
\begin{eqnarray}
& & F_{12}+F_{34}+F_{56} \pm F_{78}=0\,,\nonumber\\
& & F_{13} +F_{42}=0\,,\quad F_{57}\pm F_{86}=0\,,\quad F_{15}+F_{62}=0\,,
\nonumber\\
& & F_{14} +F_{23}=0\,,\quad F_{76}\pm F_{85}=0\,,\quad F_{16}+F_{25}=0\,,
\label{eq:2.4}\\
& & F_{73} \pm F_{48}=0\,,\quad F_{17}\pm F_{82}=0\,,\quad F_{35}+F_{64}=0\,,
\nonumber\\
& & F_{47} \pm F_{38}=0\,,\quad F_{18}\pm F_{27}=0\,,\quad F_{63}+F_{54}=0\,.
\nonumber
\end{eqnarray}

\item The 3/16 BPS equations 
\begin{eqnarray}
& & F_{12}+F_{34}=0\,,\quad F_{13}+F_{42}=0\,,\quad F_{14}+F_{23}=0\,,
\nonumber\\
& & F_{56}\pm F_{78}=0\,,\quad F_{75}\pm F_{68}=0\,,\quad F_{67}\pm F_{58}=0\,,
\nonumber\\
& & F_{15}=F_{26}=F_{37}=\pm F_{48}\,,\label{eq:2.5}\\
& & F_{16}=F_{52}=F_{47}=\pm F_{83}\,,\nonumber\\
& & F_{17}=F_{53}=F_{64}=\pm F_{28}\,,\nonumber\\
& & \pm F_{18}=F_{72}=F_{36}= F_{54}\,.\nonumber
\end{eqnarray}
Some solutions of these equations and their ADHM-like constructions
 are discussed in~\cite{cgk, popov, ohta, hio, pt, hiraoka}.

\item The 4/16 BPS equations 
\begin{eqnarray}
& & F_{12}+F_{34}=0\,,\quad F_{13}+F_{42}=0\,,\quad F_{14}+F_{23}=0\,,
\nonumber\\
& & F_{56}\pm F_{78}=0\,,\quad F_{75}\pm F_{68}=0\,,\quad 
F_{67}\pm F_{58}=0\,,\label{eq:2.6}\\
& & F_{ab}=0\quad \mathrm{for}\quad a\in \{ 1,2,3,4\}\quad 
b\in \{ 5,6,7,8\}\,.\nonumber
\end{eqnarray}
These are the subjects of the next section.

\item The 5/16 BPS equations
\begin{eqnarray}
& & F_{12}=F_{43}=F_{65}=\pm F_{78}\,,\nonumber\\
& & F_{13}=F_{24}=F_{75}=\pm F_{86}\,,\label{eq:2.7}\\
& & F_{14}=F_{32}=F_{76}=\pm F_{58}\,,\nonumber\\
& & F_{ab}=0\quad \mathrm{for}\quad a\in \{ 1,2,3,4\}\quad 
b\in \{ 5,6,7,8\}\,.\nonumber
\end{eqnarray}

\item The 6/16 BPS equations 
\begin{equation}
F_{12}=F_{43}=F_{65}=\pm F_{78}\,,\quad 
\textrm{and other components are zero} \,.\label{eq.2.8}
\end{equation}
\end{itemize}


\section{Solutions of the 1/4 BPS equations and the moduli space }

In this section, we construct some solutions of 
the 1/4 BPS equations (\ref{eq:2.6}) 
as the noncommutative instantons on $\mathbf{R}^8$ 
and interpret them as $D$-brane bound states
 with a $B$-field.
This noncommutativity is induced by a constant NS $B$-field 
on the $D8$-brane.
We also discuss some aspects of the structure of the moduli space 
of the 1/4 BPS equations in four and eight dimensions, and 
determine explicitly the metric of the moduli space 
of the $U(2)$ one-instanton.

The 1/4 BPS equations (\ref{eq:2.6}) are a copy of the 1/4 BPS equations 
on the four dimensional hyper-planes spanned by $x^1,\cdots, x^4$ and 
$x^5,\cdots, x^8$ which we call $\mathbf{R}^4$ and 
$\tilde{\mathbf{R}}^{4}$ respectively. 
Therefore we can easily construct the solutions by placing 
the four dimensional instanton solutions on each  $\mathbf{R}^4$ and 
$\tilde{\mathbf{R}}^{4}$. 
In the following, 
we consider the solutions of the gauge group $U(N)$ with 
the instanton number $k$ on the first four dimensional hyper-plane 
$\mathbf{R}^4$ and the instanton number $k'$ on the second hyper-plane
$\tilde{\mathbf{R}}^{4}$,  
and interpret them as the $D0$-$D4$-$D8$ bound states with a $B$-field.


\subsection{ADHM-like construction for the 1/4 BPS equations 
in eight dimensions}

The ADHM construction is a powerful tool to construct 
the Yang-Mills instantons in four dimensions~\cite{adhm, cg}. 
Especially, it is well-known that the instanton moduli space and the 
ADHM moduli space completely coincide. 
The ADHM construction can be extended to the 3/16 BPS equations 
in eight dimensions, which was investigated 
by several authors~\cite{cgk, ohta, hio, pt, hiraoka}. 
It is also possible to extend the ADHM construction to the 1/4 BPS equations 
on $\mathbf{R}^8$, which is studied in the following.

In order to treat the eight dimensional space, it is useful to regard 
the coordinates of $\mathbf{R}^8$ as two  
quaternionic numbers,
\begin{equation}
\mathbf{x}=\sum_{\mu =1}^8 \tilde{\sigma}_{\mu}x^{\mu}=
\left( \begin{array}{@{\,}cc@{\,}}
   z_2 & z_1\\
  -\bar{z}_1  &  \bar{z}_2
  \end{array}  \right)\,,\quad 
\mathbf{x}'=\sum_{\mu =1}^8 \tilde{\sigma}'_{\mu}x^{\mu}=
\left( \begin{array}{@{\,}cc@{\,}}
   z_4 & z_3\\
  -\bar{z}_3  &  \bar{z}_4
  \end{array}  \right)\,,\label{eq:adhm1}
\end{equation}
where we defined the eight vector matrices,
\begin{eqnarray}
\tilde{\sigma}_{\mu}&=&
\left( \begin{array}{@{\,}cccccccc@{\,}}
i\tau_1,& i\tau_2,& i\tau_3, & \textbf{1}_2,& 0, & 0, & 0, &0 
\end{array}  \right) \,,\label{eq:adhm2}\\
\tilde{\sigma}'_{\mu}&=&
\left( \begin{array}{@{\,}cccccccc@{\,}}
0, & 0, & 0, & 0, & i\tau_1, & i\tau_2, & i\tau_3, & \textbf{1}_2
\end{array}  \right) \,,\label{eq:adhm3}
\end{eqnarray}
and the four complex coordinates,
\begin{equation}
 z_1=x^2+ix^1\,,\quad z_2=x^4+ix^3\,,\quad 
z_3=x^6+ix^5\,,\quad z_4=x^8+ix^7\,.\label{eq:adhm4}
\end{equation}
Here $\tau_1,\tau_2$ and $\tau_3$ are usual Pauli matrices.

Using the $2k\times (N+2k) $ matrices $\mathbf{a}\,,\,\mathbf{b}$ 
and the $2k' \times (N+2k')$ matrices $\mathbf{a}'\,,\,\mathbf{b}'$, 
we define the Dirac-like operator,
\begin{equation}
\mathcal{D}_z = \left( \begin{array}{@{\,}cc@{\,}}
 \mathbf{a}+\mathbf{b} \mathbf{x}  &  \mathbf{0}  \\
 \mathbf{0}  &  \mathbf{a}'+\mathbf{b}'\mathbf{x}'
\end{array}  \right)\equiv 
\left( \begin{array}{@{\,}cc@{\,}}
 D_z  &  \mathbf{0}  \\
 \mathbf{0}  &  D_z^{\prime}
\end{array}  \right)\,,\label{eq:adhm5}
\end{equation}
where we also defined $D_z$ and $D_z^{\prime}$. 
We can construct the $U(N)$ gauge field as
\begin{equation}
 A_{\mu}=\mathrm{tr}\,\Psi^{\dagger}\partial_{\mu}\Psi =
\psi^{\dagger}\partial_{\mu}\psi +\psi^{\prime\dagger}\partial_{\mu}\psi'\,,
\label{eq:adhm8}
\end{equation}
where $\Psi$ is the solution of the following Dirac-like equations,
\begin{equation} 
\mathcal{D}_z \Psi= \mathcal{D}_z
\left( \begin{array}{@{\,}cc@{\,}}
 \psi & 0  \\
 0   &  \psi' 
\end{array}  \right) =0 \,,\label{eq:adhm6}
\end{equation} 
for the $(N+2k)\times N$ matrix $\psi$ and
the $(N+2k')\times N$ matrix $\psi'$, which are normalized as 
\begin{equation}
\psi^{\dagger}\psi =\mathbf{1}_{N\times N}\,,\quad
 \psi^{\prime\dagger}\psi' =\mathbf{1}_{N\times N}\,.\label{eq:adhm7}
\end{equation}
Then using the completeness equation,
\begin{equation}
\textbf{1}_{2N+2(k+k')}=\Psi\Psi^{\dagger}
+\mathcal{D}_z^{\dagger}\frac{1}{ \mathcal{D}_z\mathcal{D}^{\dagger}_z}
\mathcal{D}_z\,,\label{eq:adhm9}
\end{equation}
we can obtain the ``anti-self-dual'' gauge field strength, which 
satisfies the 1/4 BPS equations (\ref{eq:2.6}) in eight dimensions,
\begin{equation}
F_{\mu\nu} = \mathrm{tr}\,\Psi^{\dagger}\left( \partial_{\left[ \mu\right. } 
\mathcal{D}_z^{\dagger} \frac{1}{ \mathcal{D}_z\mathcal{D}_z^{\dagger}}  
\partial_{\left. \nu \right]  }\mathcal{D}_z\right)\Psi\,.\label{eq:adhm10}
\end{equation}
This can be confirmed by writing the field strength more concretely,
\begin{equation}
F_{\mu\nu}=\left\{
\begin{array}{@{\,}lll}
\psi^{\dagger}\mathbf{b}^{\dagger}
\bar{\tilde{\sigma}}_{\left[ \mu\right. }
\tilde{\sigma}_{\left. \nu \right]  } \left(D_zD_z^{\dagger}\right)^{-1}
\mathbf{b}\psi\,, & 
\textrm{for} & \mu,\nu =1,\cdots,4 \,,\\
\psi^{\prime\dagger}\mathbf{b}^{\prime\dagger}
\bar{\tilde{\sigma}}^{\prime}_{\left[ \mu\right. }
\tilde{\sigma}^{\prime}_{\left. \nu \right]  } 
\left( D'_zD_z^{\prime\dagger}\right)^{-1} \mathbf{b}^{\prime }\psi'\,, & 
\textrm{for} & \mu,\nu =5,\cdots,8 \,,\\
 0\,, &  \textrm{for} & \textrm{the other components}\,. 
\end{array}
\right.    
\label{eq:adhm12}
\end{equation}  
Here we must require that  $D_zD_z^{\dagger}$ and 
$D_z^{\prime}D_z^{\prime\dagger}$ commute with 
$\tilde{\sigma}_{\mu}$ and $\tilde{\sigma}^{\prime}_{\mu}$ respectively.
This is a necessary condition to obtain 
the ``anti-self-dual'' gauge field strength 
on $\mathbf{R}^8$. We call the equations corresponding to this condition 
the ADHM-like equations in eight dimensions 
for both the commutative and the noncommutative case.


\subsection{Noncommutative version of the ADHM-like construction for the  
1/4 BPS equations in eight dimensions}

As in the four dimensional case, 
it is also easy to extend the ADHM-like construction in eight dimensions  
to noncommutative space because of its algebraic nature.
Since we define the instantons as ``anti-self-dual'' configurations, 
the ``self-dual'' $B$-field is of interest from the viewpoint of 
the resolution of the instanton moduli space.

In this case the coordinates of $\mathbf{R}^8$ become noncommutative as
\begin{equation}
\left[ \bar{z}_1,z_1\right]=\left[ \bar{z}_2,z_2\right]=\frac{\zeta}{2},\quad
\left[ \bar{z}_3,z_3\right]=\left[ \bar{z}_4,z_4\right]=\frac{\zeta'}{2}\,,
\label{eq:nc1}
\end{equation}
for positive constant parameters $\zeta$ and $\zeta'$.
These commutation relations can be represented 
using the creation and annihilation 
operators which act on the Fock space of harmonic oscillators\,,
\begin{eqnarray}
& &  \sqrt{\frac{2}{\zeta}} z_1\,|n_1:n_2:n_3:n_4\rangle =\sqrt{ n_1+1}\,
|n_1+1:n_2:n_3:n_4\rangle\,,\nonumber\\ 
& &  \sqrt{\frac{2}{\zeta}} \bar{z}_1\,|n_1:n_2:n_3:n_4\rangle =\sqrt{ n_1}\,
|n_1-1:n_2:n_3:n_4\rangle\,,\nonumber\\
& & \sqrt{\frac{2}{\zeta}} z_2\,|n_1:n_2:n_3:n_4\rangle =\sqrt{ n_2+1}\,
|n_1:n_2+1:n_3:n_4\rangle\,,\nonumber\\
& &  \sqrt{\frac{2}{\zeta}} \bar{z}_2\,|n_1:n_2:n_3:n_4\rangle =\sqrt{ n_2}\,
|n_1:n_2-1:n_3:n_4\rangle\,,\label{eq:nc2}\\
& & \sqrt{\frac{2}{\zeta'}} z_3\,|n_1:n_2:n_3:n_4\rangle =\sqrt{ n_3+1}\,
|n_1:n_2:n_3+1:n_4\rangle\,,\nonumber\\ 
& & \sqrt{\frac{2}{\zeta'}} \bar{z}_3\,|n_1:n_2:n_3:n_4\rangle =\sqrt{ n_3}\,
|n_1:n_2:n_3-1:n_4\rangle\,,\nonumber\\
& & \sqrt{\frac{2}{\zeta'}} z_4\,|n_1:n_2:n_3:n_4\rangle =\sqrt{ n_4+1}\,
|n_1:n_2:n_3:n_4+1\rangle\,,\nonumber\\ 
& & \sqrt{\frac{2}{\zeta'}} \bar{z}_4\,|n_1:n_2:n_3:n_4\rangle =\sqrt{ n_4}\,
|n_1:n_2:n_3:n_4-1\rangle\,,\nonumber
\end{eqnarray}
where the number operators are defined by
\begin{equation}
 n_1=\frac{2}{\zeta} z_1\bar{z}_1,\quad 
 n_2=\frac{2}{\zeta} z_2\bar{z}_2,\quad
 n_3=\frac{2}{\zeta'} z_3\bar{z}_3,\quad
 n_4=\frac{2}{\zeta'} z_4\bar{z}_4\,.\label{eq:nc.3}
\end{equation}

If we also require that $D_zD_z^{\dagger}$ and 
$D_z^{\prime}D_z^{\prime\dagger}$ commutes with $\tilde{\sigma}_{\mu}$ 
and $\tilde{\sigma}_{\mu}^{\prime}$ respectively  
as in the commutative case, 
we can obtain the ``anti-self-dual'' gauge field strength on  
noncommutative $\mathbf{R}^8$. 
As in the four dimensional case, there are equivalence relations between 
different sets of the matrices $\mathbf{a},\mathbf{b},\mathbf{a}'$ 
and $\mathbf{b}'$ as
\begin{equation}
\mathbf{a}\sim M\mathbf{a} N,\quad \mathbf{b}\sim M\mathbf{b} N,\quad 
\mathbf{a}'\sim M'\mathbf{a}' N',\quad \mathbf{b}'\sim M'\mathbf{b}' N',
\end{equation}
where $M\in GL(2k,\mathbf{C})$, $N\in U(N+2k)$, $M'\in GL(2k',\mathbf{C})$ 
and $N'\in U(N+2k')$\,.
Using these relations, the
ADHM-like equations for the 1/4 BPS equations on noncommutative $\mathbf{R}^8$ 
can be obtained,
\begin{eqnarray}
\mu_{\mathbf{R}} &=& \left[ B_1,\,B_1^{\dagger} \right] +
\left[ B_2,\,B_2^{\dagger} \right] +II^{\dagger} -J^{\dagger}J=\zeta\,,
\label{eq:nc4}\\
\mu_{\mathbf{C}} &=& \left[ B_1,\,B_2 \right] + IJ =0\,,\nonumber\\
\mu_{\mathbf{R}}^{\prime} &=& \left[ B'_1,\,B_1^{\prime\dagger} \right] +
\left[ B'_2,\,B_2^{\prime\dagger} \right] +I'I^{\prime\dagger} 
-J^{\prime\dagger}J^{\prime}=\zeta^{\prime}\,,\label{eq:nc6}\\
\mu_{\mathbf{C}}^{\prime} &=& \left[ B'_1,\,B'_2 \right] + I'J' =0\,,
\nonumber
\end{eqnarray}
where $B_1,B_2$ are $k\times k$ matrices, $I,J^{\dagger}$ are $k\times N$ 
matrices, $B_1^{\prime},B_2^{\prime}$ are $k'\times k'$ matrices and 
$I^{\prime},J^{\prime\dagger}$ are $k'\times N$ matrices.
These are nothing but two sets of the four dimensional ADHM equations on 
$\mathbf{R}^4$ and $\tilde{\mathbf{R}}^{4}$.


\subsection{$U(1)$ instantons}

In this subsection, we explicitly construct the 
noncommutative $U(1)$ instanton solutions of (\ref{eq:2.6}) 
in eight dimensions.
This can be easily achieved using 
the noncommutative $U(1)$ instanton solutions in four dimensions.
Compared with the commutative case, the $U(1)$ instanton 
is already non-trivial in the noncommutative case.


\subsubsection{Four dimensional case}

Before discussing the eight dimensional case, 
we briefly review the $U(1)$ one-instanton solution in four dimensions.
Some relevant references are~\cite{ns, nekrasov, 
furuuchi, furuuchi2, furuuchi3, kly, ckt, hamanaka, popov2, tz}.

In this case, the ADHM-like equations are the same as (\ref{eq:nc4}) and
can be solved by
\begin{equation}
  B_1=B_2=J=0\,,\quad I=\sqrt{\zeta}\,.\label{u14.2}
\end{equation}
Solving the Dirac-like equations and using (\ref{eq:adhm12}), the 
anti-self-dual field strength can be obtained,
\begin{equation}
F=\frac{\zeta}{\delta (\delta +\zeta/2)(\delta +\zeta)}
\biggl\{ f_3(dz_2\wedge d\bar{z}_2 - dz_1\wedge d\bar{z}_1)
+f_{+}d\bar{z}_1\wedge dz_2 +f_{-} d\bar{z}_2\wedge dz_1 \biggr\} \,,
\label{eq:u14.3}
\end{equation}
where we defined
\begin{equation}
\delta =z_1\bar{z}_1 +z_2\bar{z}_2\,,\quad
f_3=z_1\bar{z}_1 -z_2\bar{z}_2\,,\quad
f_{+}=2z_1\bar{z}_2\,,\quad f_{-}=2z_2\bar{z}_1\,.\label{eq:u14.4}
\end{equation}
The topological charge density of (\ref{eq:u14.3}) is given by
\begin{equation}
  q=-\frac{1}{8\pi^2}\, F\wedge F 
=-\frac{\zeta^2}{\pi^2}\frac{1}{\delta (\delta +\zeta /2)^2 
(\delta +\zeta)}\,p\,,
\label{eq:u14.5}
\end{equation}
where we used
\begin{equation}
 dz^1\wedge d\bar{z}^1 \wedge dz^2\wedge d\bar{z}^2 = 
-4(\textrm{volume form})\,,\label{eq:u14.6}
\end{equation}
and defined the projection operator,
\begin{equation}
p\equiv 1-|0:0\rangle\langle 0:0| \,.\label{eq:u14.7}
\end{equation}
The topological charge of the above configuration can be calculated as
\begin{eqnarray}
  \textrm{Tr}_{\mathcal{H}}\, q &=& \left( \frac{\zeta\pi}{2}\right)^2 
\sum_{(n_1,n_2)\neq (0,0)}^{\infty} q\nonumber\\
 &=& -4 \sum_{N=1}^{\infty} \frac{1}{N(N+1)(N+2)}=-1\,, \label{eq:u14.10}
\end{eqnarray}
and coincides with the expected number $-1$\,, 
where we used the integration formula,
\begin{equation}
  \int d^4x \,\mathcal{O}(x)\to \mathrm{Tr}_{\mathcal{H}}\mathcal{O}(x)
\equiv \left( \frac{\pi\zeta}{2}\right)\sum_{(n_1,n_2)}
\langle n_1:n_2|\mathcal{O}(x)|n_1:n_2\rangle \,,\label{eq:u14.1}
\end{equation}
and the following equations,
\begin{eqnarray}
& &   \delta\,|n_1:n_2\rangle = \frac{\zeta}{2}(n_1+n_2)|n_1:n_2\rangle \,,
\label{eq:u14.8}\\
& &  \sum_{(n_1,n_2)\neq (0,0)}^{\infty} \langle N|\mathcal{O}(\delta)|N
\rangle =\sum_{N=1}^{\infty}(N+1)\langle N|\mathcal{O}(\delta)|N\rangle\,.
\label{eq:u14.9}
\end{eqnarray}
This fact enables us to interpret the noncommutative $U(1)$ one-instanton 
as the bound state of the $D0$-brane and the $D4$-brane with a $B$-field.


\subsubsection{Eight dimensional case}

If we place the noncommutative instantons in each four dimensional 
hyper-plane $\mathbf{R}^4$ and $\tilde{\mathbf{R}}^4$, 
we can easily construct the eight dimensional solutions.


\begin{flushleft}
\textbf{One-instanton solution}
\end{flushleft}

First we consider the simplest case where $N=k=k'=1$.
In this case, the ADHM-like equations (\ref{eq:nc4})(\ref{eq:nc6}) 
can be solved by
\begin{eqnarray}
& &   B_1=B_2=J=0\,,\quad I=\sqrt{\zeta}\,,\label{eq:u18.1}\\
& &   B'_1=B'_2=J'=0\,,\quad I'=\sqrt{\zeta'}\,.\label{eq:u18.2}
\end{eqnarray}
As in the four dimensional case, the anti-self-dual
field strength can be obtained as 
\begin{equation}
 F = f+\tilde{f}\,,
\end{equation}
where we defined $f$ and $\tilde{f}$ by
\begin{eqnarray}
f &=&  \frac{\zeta}{\delta (\delta +\zeta/2)(\delta +\zeta)}
\biggl\{ f_3(dz_2\wedge d\bar{z}_2 - dz_1\wedge d\bar{z}_1)
+f_{+}d\bar{z}_1\wedge dz_2 +f_{-} d\bar{z}_2\wedge dz_1 \biggr\} 
\,,\nonumber\\
\tilde{f} &=& \frac{\zeta'}{\delta' (\delta' +\zeta'/2)(\delta' +\zeta')}
\biggl\{ f_3^{\prime}(dz_4\wedge d\bar{z}_4 - dz_3\wedge d\bar{z}_3)
+f_{+}^{\prime} d\bar{z}_3\wedge dz_4 +f_{-}^{\prime}
 d\bar{z}_4\wedge dz_3 \biggr\}\,.\nonumber
\end{eqnarray}
We also defined
\begin{eqnarray}
& & \delta =z_1\bar{z}_1 +z_2\bar{z}_2\,,\quad 
f_3=z_1\bar{z}_1 -z_2\bar{z}_2\,,\quad
f_{+}=2z_1\bar{z}_2\,,\quad f_{-}=2z_2\bar{z}_1 \,,\label{eq:u18.4}\\
& & \delta^{\prime} =z_3\bar{z}_3 +z_4\bar{z}_4\,,\quad 
f_3^{\prime}=z_3\bar{z}_3 -z_4\bar{z}_4\,,\quad
f_{+}^{\prime}=2z_3\bar{z}_4\,,\quad f_{-}^{\prime}=2z_4\bar{z}_3 \,.
\label{eq:u18.5}
\end{eqnarray}

This configuration has the four-form charges over the corresponding 
four dimensional hyper-plane,
\begin{equation}
 k= -\frac{1}{2!(2\pi)^2 }\int_{\mathbf{R}^4} d^4x\, f\wedge f=-1,\quad
k'= -\frac{1}{2!(2\pi)^2 }\int_{\mathbf{\tilde{R}}^4} d^4x\, \tilde{f}
\wedge \tilde{f}=-1\,.
\label{eq:u18.6}
\end{equation}
These charges can be interpreted as the charge of the $D4$-brane 
bound to the $D8$-brane.
This configuration also has the eight-form charge,
\begin{equation}
Q=\frac{1}{4!(2\pi)^4 }\int_{\mathbf{R}^4\times \tilde{\mathbf{R}}^4} 
d^8x \,F\wedge F\wedge F\wedge F\,,\label{eq:u18.7}
\end{equation}
which can be calculated explicitly as
\begin{equation}
Q = \frac{6\cdot 16}{4!(2\pi)^4} 
\left( \frac{\zeta\pi}{2}\right)^2 \left( \frac{\zeta'\pi}{2}\right)^2 
\sum \biggl\{ \frac{-2\zeta^2}{\delta (\delta +\zeta /2)^2
(\delta +\zeta)}\biggr\} 
 \biggl\{ \frac{-2\zeta^{\prime 2}}{\delta' (\delta' +\zeta^{\prime} /2)^2
(\delta' +\zeta^{\prime})}\biggr\} =1\,,\label{eq:u18.8}
\end{equation}
where we used
\begin{equation}
 dz^1\wedge d\bar{z}^1 \wedge dz^2\wedge d\bar{z}^2
 \wedge dz^3\wedge d\bar{z}^3  \wedge dz^4\wedge d\bar{z}^4= 
16(\textrm{volume form})\,.\label{eq:u18.9}
\end{equation}
This charge can be interpreted as the $D0$-brane charge. 
As a result, we can regard the above configuration as 
the bound state of the $D0$-$D4$-$D8$ with a $B$-field.


\begin{flushleft}
\textbf{Multi-instanton solutions}
\end{flushleft}

The above construction can be easily generalized 
to the multi-instanton solutions.
If the solution has the $D4$-brane charge $k$ for the first four dimensional 
hyper-plane $\mathbf{R}^4$ and $k'$ for the second one $\tilde{\mathbf{R}}^4$, 
this configuration also has the $D0$-brane charge,
\begin{equation}
  Q=kk'\,.\label{eq:u18.10}
\end{equation}
Therefore this configuration can be interpreted as the bound states 
of the $D0$-$D4$-$D8$ with a $B$-field.


\subsection{Structure of the moduli space of the $U(2)$ one-instanton}

Similar constructions are also possible for the case of the noncommutative 
$U(2)$ instanton. But since the constructions are straightforward,  
we give the results in the case of the $U(2)$ one-instanton in the appendix. 

In this subsection, we consider the structure of the moduli space 
of the noncommutative $U(2)$ instanton solutions. 
Though the noncommutative $U(1)$ one-instanton 
has no degrees of freedom except its position, 
the $U(2)$ one-instanton is the first example which has the non-trivial 
structure of the moduli space. 
In the following, we mainly focus on this case.

\begin{flushleft}
\textbf{Four dimensional case}
\end{flushleft}

To begin with, we describe the procedure to obtain the moduli space metric 
from the ADHM data. Noncommutativity parameter $\zeta$ deforms the metric 
from that in the commutative case.
In four dimensions, it is well-known that 
the moduli space of the noncommutative $U(N)$ 
$k$-instanton is given by the hyper-K\"{a}hler 
quotient~\cite{adhm, lty, osborn, kly2},
\[ \widetilde{\mathcal{M}}(k,N)=\left\{ \mu_{\mathbf{R}}^{-1}(\zeta)\cap 
\mu_{\mathbf{C}}^{-1}(0) \right\} /U(k)\,, \]
where the moment maps $\mu_{\mathbf{R}}$ and $\mu_{\mathbf{C}}$ are defined 
as follows,
\begin{eqnarray}
\mu_{\mathbf{R}} &=& \left[ B_1,\,B_1^{\dagger} \right] +
\left[ B_2,\,B_2^{\dagger} \right] +II^{\dagger} -J^{\dagger}J\,,
\label{eq:u2.1}\\
\mu_{\mathbf{C}} &=& \left[ B_1,\,B_2 \right] + IJ\,.\nonumber
\end{eqnarray}
Here, $B_1, B_2$ are $k\times k$ matrices and $I,\,J^{\dagger}$
 are $k\times N$ matrices. 
When $\zeta =0$ i.e. the commutative case, 
this moduli space of dimension $4Nk$ is singular when the instantons shrink 
to zero size.

The tangent vectors for the ADHM data are a sum of the differentials 
with the independent parameters and the infinitesimal gauge transformation,
\begin{eqnarray}
\delta B_1 &=& dB_1 -i\left[ \alpha,\,B_1\right]\,,\nonumber\\
\delta B_2 &=& dB_2 -i\left[ \alpha,\,B_2\right]\,,\label{eq:u2.3}\\
\delta I &=& dI -i\alpha I \,,\nonumber\\
\delta J^{\dagger} &=& dJ^{\dagger} -i\alpha J^{\dagger}\,,\nonumber
\end{eqnarray}
where $\alpha$ is a hermitian matrix and the differential $d$ acts 
on all parameters. These tangent vectors must satisfy the 
linearized ADHM equations,
\begin{eqnarray}
& & \left[ \delta B_1,\,B_1^{\dagger}\right] 
+\left[ B_1,\,\delta B_1^{\dagger}\right]
 +\left[ \delta B_2,\,B_2^{\dagger}\right] 
+\left[ B_2,\,\delta B_2^{\dagger}\right] +
 \delta II^{\dagger} +I\delta I^{\dagger} -\delta J^{\dagger} J 
-J^{\dagger}\delta J =0\,, \nonumber\\
& & \left[ \delta B_1,B_2\right] +\left[ B_1,\delta B_2\right] 
+\delta IJ+I\delta J =0\,,\label{eq:u2.4}
\end{eqnarray}
and the linearized 
background gauge fixing condition,
\begin{equation}
\left[ \delta B_1,\,B_1^{\dagger}\right] 
-\left[ B_1,\,\delta B_1^{\dagger}\right]
 +\left[ \delta B_2,\,B_2^{\dagger}\right] 
-\left[ B_2,\,\delta B_2^{\dagger}\right] +
 \delta II^{\dagger} -I\delta I^{\dagger} +\delta J^{\dagger} J 
-J^{\dagger}\delta J =0\,. \label{eq:u2.5}
\end{equation}
From the above equations, we can obtain
\begin{equation}
\left[ \delta B_1,\,B_1^{\dagger}\right] +
\left[ \delta B_2,\,B_2^{\dagger}\right] +\delta II^{\dagger}
-J^{\dagger} \delta J=0\,,\label{eq:u2.6}
\end{equation}
which fixes the matrix $\alpha$.
Finally the moduli space metric can be obtained from the formula,
\begin{equation}
ds^2= \mathrm{tr}\left( \delta B_1 \delta B_1^{\dagger} 
+ \delta B_2 \delta B_2^{\dagger}
+ \delta I\delta I^{\dagger} +\delta J^{\dagger}\delta J\right) \,.
\label{eq:u2.7}
\end{equation}

In the following, we explicitly follow the above procedure 
for the case of the $U(2)$ one-instanton.
Now $B_1$ and $B_2$ are numbers, the ADHM equations reduce to
\begin{equation}
 II^{\dagger} -J^{\dagger}J=\zeta \,,\quad  IJ =0\,.\label{eq:u2.9}
\end{equation}
The general solution to these equations is given by
\begin{equation}
 B_1=c_1,\quad B_2=c_2,\quad I =\left(
\begin{array}{@{\,}cc@{\,}}
  w_1 & w_2 
\end{array} \right),\quad 
 J^{\dagger} = B
\left(
\begin{array}{@{\,}cc@{\,}}
  -\bar{w}_2  & \bar{w}_1
\end{array} \right)\,,\label{eq:u2.10}
\end{equation}
where $c_{1,2}$ and $w_{1,2}$ are regarded as independent variables 
which parameterize the eight dimensional manifold 
$\widetilde{\mathcal{M}}(1,2)$, and we defined $B$ by
\begin{equation}
 B\equiv \sqrt{ 1-\frac{\zeta}{A}},\quad \mathrm{where}\quad 
 A\equiv |w_1|^2 +|w_2|^2\geq \zeta \,.\label{eq:u2.11}
\end{equation}
Here we have fixed the global $U(1)$ symmetry.  
As we will see, $c_1$ and $c_2$ can be interpreted as the position of
 the instanton on $\mathbf{R}^4$, and 
$w_1$ and $w_2$ are the degrees of freedom of the instanton size and 
the global $SU(2)$ transformations as discussed in~\cite{kly2}.  
Since we can make the following identification,
\begin{equation}
 w_1\cong -w_1,\quad w_2\cong -w_2\,,\label{eq:u2.12}
\end{equation}
the metric of the relative moduli space which is the metric 
at the center of mass frame is expected to be the Eguchi-Hanson metric.
This is also expected from the fact that the relative moduli space of the  
noncommutative $U(1)$ two-instanton is the Eguchi-Hanson space~\cite{lty} and 
that the dimensions of $\widetilde{\mathcal{M}}(1,2)$ 
and $\widetilde{\mathcal{M}}(2,1)$ are the same\cite{dhk}.

Now $\delta B_1 =dB_1=dc_1$ and $\delta B_2 =dB_2 =dc_2$ 
are nothing but numbers, we must solve (\ref{eq:u2.6})\,,
\begin{equation}
\delta II^{\dagger}= J^{\dagger} \delta J\,.\label{eq:u2.13}
\end{equation}
Then we can determine $\alpha$ as
\begin{equation}
\alpha =-i\frac{\zeta}{2A(2A-\zeta)} ( \bar{w}_idw_i-w_id\bar{w}_i)\,,
\label{eq:u2.17}
\end{equation}
where we defined the differential of $A$ as
\begin{eqnarray}
& & dA =\partial A +\bar{\partial} A\,, \label{eq:u2.15}\\
& & \partial A=\bar{w}_1 dw_1 +\bar{w}_2dw_2 =\bar{w}_i dw_i\,,
\quad \bar{\partial} A=w_1 d\bar{w}_1 +w_2d\bar{w}_2 =w_i d\bar{w}_i\,.
\nonumber
\end{eqnarray}

Since 
\begin{equation}
\delta B_1 \delta B_1^{\dagger} + \delta B_2 \delta B_2^{\dagger}
=dc_1d\bar{c}_1 +dc_2d\bar{c}_2\label{eq:u2.18}
\end{equation}
gives the metric of the coordinates of the center of mass, 
the metric of the relative moduli space  
can be obtained substituting (\ref{eq:u2.17}) into $\alpha$,
\begin{eqnarray}
\delta I\delta I^{\dagger} +\delta J^{\dagger}\delta J &=&
(B^2+1)dw_id\bar{w}_i +\left( \frac{\zeta^2}{4A^3B^2} +\frac{\zeta}{2A^2}
\right) (\bar{w}_idw_i +w_id\bar{w}_i)^2 \nonumber\\
& & \quad +\frac{\xi^2}{4A^2(2A-\zeta)} (\bar{w}_idw_i -w_id\bar{w}_i)^2 \,.
\label{eq:u2.20}
\end{eqnarray}
We can parameterize $w_1$ and $w_2$ using the Euler angles as
\begin{eqnarray}
w_1 &=& r\cos \left(\frac{\theta}{2}\right)\,\exp 
\biggl\{\frac{i}{2}(\psi +\varphi)\biggr\}\,,
\label{eq:u2.21}\\
w_2 &=& r\sin \left( \frac{\theta}{2}\right)\,\exp 
\biggl\{\frac{i}{2}(\psi -\varphi)\biggr\}\,,
\label{eq:u2.22}
\end{eqnarray}
where the ranges of the angular variables are given by
\begin{equation}
  0\leq \theta\leq \pi,\quad 0\leq \varphi\leq 2\pi,\quad 
0\leq \psi\leq 2\pi\,.
\end{equation} 
In these variables, the metric of the relative moduli space 
$ds_{\mathrm{rel}}^2 = \delta I\delta I^{\dagger}
 +\delta J^{\dagger}\delta J$ becomes
\begin{equation}
ds_{\mathrm{rel}}^2 = \left( 2-\frac{\zeta}{r^2}\right)\left\{ 
\frac{dr^2}{1-\zeta /r^2} +\frac{r^2}{4}(\sigma_1^2 +\sigma_2^2)
+\frac{r^4(r^2-\zeta)}{(2r^2-\zeta)^2}\sigma_3^2\right\}\,,\label{eq:u2.25}
\end{equation}
where we used the $SU(2)_L$ invariant one-forms,
\begin{eqnarray}
\sigma_1 &=& -\sin \psi d\theta +\cos\psi\sin\theta d\varphi\,,\nonumber\\
\sigma_2 &=& \cos \psi d\theta +\sin\psi\sin\theta d\varphi\,,
\label{eq:u2.23}\\
\sigma_3 &=& d\psi +\cos\theta d\varphi\,,\nonumber
\end{eqnarray}
which satisfy the $SU(2)$ Mauer-Cartan equation,
\begin{equation}
d\sigma_i =\frac{1}{2}\epsilon_{ijk}\sigma_j\wedge\sigma_k\,.\label{eq:u2.24}
\end{equation}
Near the origin, the metric of the relative moduli space becomes the one  
for $\mathbf{R}^2\times \mathbf{S}^2$ which is nonsingular.
This gives the explicit realization of the resolution of 
the instanton moduli space~\cite{ns}.
If we further change the variable as
\begin{equation}
  u^2\equiv 2r^2-\zeta\geq \zeta\,,\label{eq:u2.26}
\end{equation}
then we can get the standard Eguchi-Hanson metric, 
which is hyper-K\"{a}hler and has the holonomy $Sp(1)$,
\begin{equation}
 ds_{\mathrm{rel}}^2=ds_{\mathrm{EH}}^2= 
\frac{du^2}{1-\zeta^2 /u^4} +\frac{u^2}{4} \left\{ \sigma_1^2 +\sigma_2^2
+\left( 1-\frac{\zeta^2}{u^4}\right)\sigma_3^2\right\}\,.\label{eq:u2.27}
\end{equation}
This agrees with the result in~\cite{dhk, ly, kly4}.


\newpage

\begin{flushleft}
\textbf{Eight dimensional case}
\end{flushleft}

Since the 1/4 BPS equations (\ref{eq:2.6}) and 
the ADHM-like equations for them (\ref{eq:nc4})(\ref{eq:nc6}) 
in the eight dimensional case 
are nothing but two sets of the four dimensional ones, 
we can expect that the relative moduli space in the eight dimensional case 
becomes the direct product of the relative moduli spaces in
the four dimensional case. 
In particular, the metric of the moduli space of the $U(2)$ one-instanton  
in the eight dimensional case becomes 
\begin{equation}
 ds_{\mathrm{rel}}^2=ds_{\mathrm{EH}}^2 
+d\tilde{s}^2_{\widetilde{\mathrm{EH}}}\,, \label{eq:u2.28}
\end{equation} 
with the holonomy $Sp(1) \times Sp(1)$, 
where $ds_{\mathrm{EH}}^2$ is the metric of the relative moduli 
space of the $U(2)$ 
one-instanton on noncommutative $\mathbf{R}^4$ 
and $d\tilde{s}_{\widetilde{\mathrm{EH}}}^2$ 
is the similar metric  on $\tilde{\mathbf{R}}^4$.

The moduli space metric of the $U(N)$ one-instanton on noncommutative 
$\mathbf{R}^4$ in its center of mass frame was found to be the 
$4N-4$ dimensional hyper-K\"{a}hler Calabi metric $ds_{\mathrm{Calabi}}^2$ 
in~\cite{ly} 
with one scale parameter and the principal orbit $SU(N+1)/U(N-1)$. 
The authors of~\cite{ly} studied the noncommutative caloron solution 
on $\mathbf{R}^3\times S^1$ 
and took the large radius limit of $S^1$.
Although the relations between the ADHM data and the variables in the metric 
are not yet made clear, it was pointed out in~\cite{kly4} that 
the above metric becomes the hyper-K\"{a}hler Calabi metric 
$ds_{\mathrm{Calabi}}^2$ 
constructed by~\cite{cglp}.
The only exception is the $N=2$ case where we have explicitly 
written down the relations as 
(\ref{eq:u2.10}), (\ref{eq:u2.21}) and (\ref{eq:u2.22})\,.
Having this result at hand, it is easy to extend (\ref{eq:u2.28}) to
\begin{equation}
 ds_{\mathrm{rel}}^2=ds_{\mathrm{Calabi}}^2 
+d\tilde{s}^2_{\widetilde{\mathrm{Calabi}}}\,, \label{eq:u2.29}
\end{equation} 
where $ds_{\mathrm{Calabi}}^2$ is the metric of the relative moduli space  
of the $U(N)$ one-instanton on noncommutative $\mathbf{R}^4$ 
and $d\tilde{s}_{\widetilde{\mathrm{Calabi}}}^2$ is 
the similar metric on noncommutative $\tilde{\mathbf{R}}^4$.


\section{Discussions and Comments}

In this paper, we focused on the 1/4 BPS equations  
of the noncommutative Yang-Mills theory 
in eight dimensions, found by~\cite{blp}. 
We constructed some explicit solutions by means of 
the noncommutative version of the ADHM-like construction in eight dimensions, 
and showed that our solutions can be interpreted 
as the bound states of the $D0$-$D4$-$D8$ with a $B$-field
from the calculation of their topological charges.
We also discussed the structure of the moduli space of the 1/4 BPS equations 
and determined the metric of the moduli space of the $U(2)$ one-instanton 
in four and eight dimensions.

There is however room for the generalization 
of our ADHM-like construction. 
It can only produce the field strength $F_{ab}\,(a,b=1,\cdots, 4)$ 
which depends on the coordinates $x^1,\cdots ,x^4$, and 
$F_{ab}\,(a,b=5,\cdots 8)$ 
which depends on the coordinates $x^5,\cdots ,x^8$.
Therefore we should seek for the general ADHM-like construction applicable 
to more complicated configurations. 
And it should also be checked whether the instanton number is the 
expected one.

We point out the relation between our ADHM-like construction for 
the 1/4 BPS solutions  
and that for the 3/16 BPS solutions 
which was used in~\cite{cgk, ohta, pt, hiraoka}.
It is of interest that when $k=k^{\prime}$ 
we can obtain the Dirac-like operator $D_z +D_z^{\prime}$ 
for the construction of the 3/16 BPS configurations 
from our Dirac-like operator (\ref{eq:adhm5}). 
In this construction, the field strength contains the 
``self-dual'' tensor,
\[
  \overline{N}_{\mu\nu}=\frac{1}{2}\left( \Sigma_{\mu}\Sigma_{\nu}^{\dagger}
-\Sigma_{\nu}\Sigma_{\mu}^{\dagger}\right)\,,
\]
with the definition,
\[
\Sigma_{\mu}\equiv \left( \begin{array}{@{\,}c@{\,}}
  \tilde{\sigma}_{\mu} \\
  \tilde{\sigma}_{\mu}^{\prime}   
\end{array}  \right)  \,,
\] 
which satisfies the 3/16 BPS equations (\ref{eq:2.5})\,. 
(Strictly speaking, we have to take different choices of the signatures of 
$\alpha$s in the notation of~\cite{blp}.) 
This fact will help extend the noncommutative version of the 
ADHM-like equations (\ref{eq:nc4})(\ref{eq:nc6}) 
for the 1/4 BPS solutions to that for the 3/16 BPS solutions 
in a systematic way. 
The tentative ADHM-like equations for the 3/16 BPS solutions in 
noncommutative $\mathbf{R}^8$ have been proposed in~\cite{ohta}. 
From these considerations,  
the holonomy group $Sp(1)\times Sp(1)$ obtained in the last subsection 
can be regarded as the subgroup of the holonomy group  
$Sp(2)$ of the 3/16 BPS solutions in eight dimensions.


\bigskip
\bigskip
\centerline{\bf Acknowledgments}

\vskip 0.6cm

We would like to thank H. Ishikawa and S. Watamura
 for useful comments, reading manuscripts and encouragements.


\appendix
\renewcommand{\thesection}%
             {\Alph{section}}
\addtocontents{toc}%
   {\protect\vspace{5mm}}
\addtocontents{toc}%
   {\protect\leftline%
       {\large\bfseries Appendix}}

\section{$U(2)$ one-instanton solution in eight dimensions}

In this appendix, 
we give the results in the case of the $U(2)$ one-instanton solution 
in eight dimensions. 
The ADHM-like equations (\ref{eq:nc4})(\ref{eq:nc6}) can be solved by
\begin{eqnarray}
& & B_1 =B_2 =0,\quad I=
\left( \begin{array}{@{\,}cc@{\,}}
 \sqrt{ \rho^2 +\zeta} & 0   
\end{array}  \right),\quad J^{\dagger}=
\left( \begin{array}{@{\,}cc@{\,}}
 0 &  \rho    
\end{array}  \right),\\
& & B_1^{\prime} =B_2^{\prime} =0,\quad I^{\prime}=
\left( \begin{array}{@{\,}cc@{\,}}
 \sqrt{ \rho^{\prime 2} +\zeta^{\prime}} & 0   
\end{array}  \right),\quad J^{\prime\dagger}=
\left( \begin{array}{@{\,}cc@{\,}}
 0 &  \rho^{\prime}    
\end{array}  \right),
\end{eqnarray}
where $\rho$ and $\rho^{\prime}$ parameterize the classical size of the 
instantons on $\mathbf{R}^4$ and $\tilde{\mathbf{R}}^{4}$ respectively.

The strength of the gauge field can be explicitly calculated as
\begin{equation}
F = f+\tilde{f}\,,\label{eq:a1}
\end{equation}
where $f$ and $\tilde{f}$ are defined as follows,
\begin{eqnarray}
f &=& \left( \begin{array}{@{\,}cc@{\,}}
 \frac{1}{2} B_1 (z_2\bar{z}_2 -z_1\bar{z}_1) & B_2 (z_1z_2)\\
  B_3 (\bar{z}_1\bar{z}_2) & \frac{1}{2} B_4  (z_1\bar{z}_1 -z_2\bar{z}_2)  
\end{array}  \right)\,
(d\bar{z}_1\wedge dz_1 -d\bar{z}_2 \wedge dz_2) \nonumber\\
 & & \quad + \left( \begin{array}{@{\,}cc@{\,}}
  -B_1 (z_1\bar{z}_2)  & -B_2 (z_1z_1)\\
  B_3 (\bar{z}_2\bar{z}_2) &  B_4  (z_1\bar{z}_2)  
\end{array}  \right)\,
d\bar{z}_1\wedge dz_2 + 
\left( \begin{array}{@{\,}cc@{\,}}
  B_1 (\bar{z}_1 z_2)  & -B_2 (z_2z_2)\\
  B_3 (\bar{z}_1\bar{z}_1) &  -B_4  (\bar{z}_1 z_2)  
\end{array}  \right)\,
dz_1\wedge d\bar{z}_2\,,\nonumber\\
\tilde{f} &=& \left( \begin{array}{@{\,}cc@{\,}}
 \frac{1}{2} B'_1 (z_4\bar{z}_4 -z_3\bar{z}_3) & B'_2 (z_3z_4)\\
  B'_3 (\bar{z}_3\bar{z}_4) & \frac{1}{2} B'_4  (z_3\bar{z}_3 -z_4\bar{z}_4)  
\end{array}  \right)\,
(d\bar{z}_3\wedge dz_3 -d\bar{z}_4 \wedge dz_4) \nonumber\\
 & & \quad + \left( \begin{array}{@{\,}cc@{\,}}
  -B'_1 (z_3\bar{z}_4)  & -B'_2 (z_3z_3)\\
  B'_3 (\bar{z}_4\bar{z}_4) &  B'_4  (z_3\bar{z}_4)  
\end{array}  \right)\,
d\bar{z}_3\wedge dz_4 + 
\left( \begin{array}{@{\,}cc@{\,}}
  B'_1 (\bar{z}_3 z_4)  & -B'_2 (z_4z_4)\\
  B'_3 (\bar{z}_3\bar{z}_3) &  -B'_4  (\bar{z}_3 z_4)  
\end{array}  \right)\,
dz_3\wedge d\bar{z}_4\,.\nonumber
\end{eqnarray}
We also defined the following equations, 
\begin{eqnarray}
B_1 &=& \frac{ -2(\rho^2 +\zeta)}{\delta (\delta +\rho^2 +\zeta/2)
(\delta +\rho^2 +\zeta)}\,,\label{eq:a2}\\
B_2 &=& \frac{ 2\rho\sqrt{\rho^2 +\zeta}}
{ \delta (\delta +\rho^2 +\zeta/2)(\delta +\rho^2 +\zeta)}
\sqrt{ \frac{ \delta +\rho^2 +\zeta}{ \delta +\rho^2}}\,,\label{eq:a3}\\
B_3 &=& \frac{ 2\rho\sqrt{\rho^2 +\zeta}}
{ (\delta +\zeta) (\delta +\rho^2 +\zeta)(\delta +\rho^2 +3\zeta /2)}
\sqrt{ \frac{ \delta +\rho^2 +\zeta}{ \delta +\rho^2 +2\zeta }}\,,
\label{eq:a4}\\
B_4 &=& \frac{-2\rho^2}{(\delta +\zeta) (\delta +\rho^2 +\zeta)
(\delta +\rho^2 +3\zeta /2)}\,,\label{eq:a4}
\end{eqnarray}
and $B_1^{\prime},\cdots, B_4^{\prime}$ are defined similarly 
by replacing $\delta, \zeta$ and $\rho$ with 
$\delta', \zeta'$ and $\rho'$ respectively.
It can be easily checked that the above field strength (\ref{eq:a1}) 
satisfies the 1/4 BPS equations (\ref{eq:2.6}) in eight dimensions. 

The calculations of the four-form charge (\ref{eq:u18.6}) 
are the same as that of~\cite{kly}, 
and this shows that our solution (\ref{eq:a1}) carries the 
$D4$-brane charges. 
The eight-form charge $Q$ (\ref{eq:u18.7}) 
of our solution (\ref{eq:a1}) can be calculated only numerically. 
But since it is believed that the topological charge does not depend on the 
parameters of size $\rho$ and $\rho^{\prime}$, 
the eight-form charge can be calculated in the limit of $\rho,\,\rho' \to 0$ . 
Since the part of the $U(1)$ one-instanton only contributes in this limit, 
the eight-form charge can be calculated 
as in the case of the $U(1)$ one-instanton. 
The result is $Q=1$. From these facts, we can regard the configuration  
 (\ref{eq:a1}) as the bound states of the $D0$-$D4$-$D8$ with a $B$-field.


\newpage

\bibliography{}

\end{document}